# OPEN SOURCE SOFTWARE FOR DIGITAL PRESERVATION REPOSITORIES: A SURVEY

Carlos André Rosa[1], Olga Craveiro[1,2] and Patricio Domingues[1,3]

[1]School of Technology and Management, Polytechnic Institute of Leiria, Portugal
[2]CISUC, Universityof Coimbra, PortugalandAlgoritmi, Universityof Minho, Portugal
[3]Instituto de Telecomunicações, Portugal

*ABSTRACT*

*In the digital age, the amount of data produced is growing exponentially. Governments and institutions can no longer rely on old methods for storing data and passing on the knowledge to future generations. Digital data preservation is a mandatory issue that needs proper strategies and tools. With this awareness, efforts are being made to create and perfect software solutions capable of responding to the challenge of properly preserving digital information. This paper focuses on the state-of-the-art in open-source software solutions for the digital preservation and curation field used to assimilate and disseminate information to designated audiences. Eleven open source projects for digital preservation are surveyed in areas such as supported standards and protocols, strategies for preservation, methodologies for reporting, dynamic of development, targeted operating systems, multilingual support and open source license. Furthermore, five of these open source projects, are further analysed, with focus on features deemed important for the area. Along open source solutions, the paper also briefly surveys the standards and protocols relevant for digital data preservation. The area of digital data preservation repositories has several open source solutions, which can form the base to overcome the challenges to reach mature and reliable digital data preservation.*

.*KEYWORDS*

*Digital data, preservation, repositories, open source*

## 1. INTRODUCTION

Information preservation can simply be defined as the set of processes to store, index and access information[1]. In recent years, the creation of digital content has grown exponentially. Gantz and Reinsel report that the so called digital universe will grow from 2005 to 2020 by a factor of 300, from 130 exabytes to 40,000 exabytes[2]. They also predict that the whole set of data will double roughly every two years from 2012 to 2020. Digital video is a good example of the current data deluge: the demand for increasing resolutions and higher frame rates, despite all improvements in compression, have substantially increased the size of video files. Smartphones, with all their data sensors, namely photo and video recording capabilities, are also major contributors to the current massive production of data [3]. The Internet of Things (IoT) is poised to generate increasing amount of data, even if IoT middleware can help by reducing the volume of data to store and preserve [4]. The sheer volume of digital information to preserve is immense and will continue to grow over the years. In fact, major trends like Big Data have fostered the perception of digital data as valuable assets, strengthening the need for digital data preservation and henceforth for proper digital repositories [5]. This way, the field of digital information preservation has to address a huge challenge.

The panorama of information preservation has substantially changed with the advent of the digital era. In fact, when paper was the main medium globally adopted for storing knowledge and information, libraries were the obvious and natural places responsible for guarding, protecting and maintaining all information stored in printed formats, such as, books, papers or other. It was not until 1960's that archivists and librarians felt concerned about the preservation of electronic records. This fostered the emergence of the Machine Readable Records branch, formed at the National Archives in the USA [6].

In the early 1990's, materials started being ported from printed formats to digital ones, being "re-born" digital. This rose awareness for the need to address preservation to digital-only data. Indeed, if exposed to mild humidity conditions and kept in a moderate environment, paper is relatively easy to preserve, or at least has a lifetime measured in decades and thus preservation can be organized accordingly. In addition, techniques such as microfilming allowed for affordable and durable preservation of paper-based documents [7]. The same does not apply to digital formats. Even if the base of digital formats is simple binary 0 and 1, digital encompasses a rich set of various resources such as text, audio/video, images, computer programs, just to name a few. Besides the main data, additional information regarding the resources – format, software environment, operating systems, etc. – is required to properly preserve and access digital data. Digital formats are very fragile and even on controlled environments, there must be an active management to assure their good shape and longevity [6][7].

The paper paradigm shift to the digital reality clearly reflected itself on other knowledge institutions, not only libraries and archives. Schools, universities and other institutions also found themselves in a situation where using paper as the main support for storing information was no longer the best choice, either because of storage space issues, preservation issues or simply because of the advance of technology.It no longer made sense to keep using outdated and less flexible means to keep information. However, if on one side there was already an awareness about the need to preserve digital information, on the other side, data repositories that followed or implemented those standards were scarce or inexistent. The logic step for these institutions was to develop their own solutions and implementations of digital preservation repositories. Their own premises and academic communities were the perfect audience to test and perfect them. Much of the software featured in this survey has its root from an academic reality.

Today, we are facing yet another challenge: the Internet. In a world where the demand for being always connected is higher than ever before, the global network and its omnipresent nature make it the obvious choice to store and disseminate information and knowledge. With an exponential growth observed during the 1990's, the volume of information available on the Internet expanded as well. However, unlike printed formats, its ephemeral existence and highly volatile availability were shortly noticed. The very nature of the Internet makes it the perfect place to publish information that frequently is not available elsewhere. The awareness and need to ensure the preservation and long-term access to this information gave birth to web archiving [8], an important subset of digital information preservation. Indeed, consisting in the collection of information available in the World Wide Web for future access, the process of harvesting that information is challenging due to its heterogeneous nature. For this purpose, the WARC standard [9] was created in 2009 by the International Internet Preservation Consortium (IIPC). Standing for *Web ARChive*, WARC specifies a method for combining multiple digital resources into an aggregate archival file together with related information to be used by web crawling software when harvesting information from websites[10]. The resulting WARC files are passible of being ingested and stored on digital repositories for preservation. However, according to a recent survey [10], very few institutions are effectively downloading the WARC files generated by the web crawlers and storing them in local preservation systems or repositories. In spite of not being a common practice among institutions, there is a recognition for the need to perform preservation of web content.

The main motivation for this paper is to help to fill a void: to the best of our knowledge, no up-to-date survey exists for open source digital preservation software. From our research, the most recent one isdated November 2010[11]. Furthermore, none of the other studies focuses exclusively on digital repositories software (e.g. [12] and [13]). Indeed, much of the scientific literature focuses on general purpose repositories. Instead, our work targets open source software repositories for digital information preservation. While they share some common requirements and properties, the two types of repositories are quite different. General-purpose repositories aim to ingest data and ensure means to store and make accessible the ingested data. On the other hand, a digital preservation repository needs to implementat least six high level services as defined by the Open Archive Information System (OAIS) reference model[14], as we shall see later in Section 2.1.

The main contribution of this paper is to present a wide-scale comparisons of leading open source software solutions that can appropriately store and preserve digital information. The paper highlights the main features of each system, the licensing model, the main preservation capabilities and which standards and protocols are supported. Furthermore, the survey provides an in-depth analysis of five of the most relevant open source solutions for digital information preservation. This way, the paper aims to provide a reference for anyone who aims to build a preservation-enabled digital library to make an informed choice. We believe that the paper is of interest even to potential clients of fee-based solutions, who can further compare their targeted commercial solutions with open source software ones.

This document introduces the subject of digital information preservation by giving an historical framing,pointing out the reasons why institutions and organizations are concerned about the preservation of their digital assets. In section 2, the paper reviews the main models, standards and protocols for digital information preservation. Section 3 compares eleven open source solutions for digital information preservation. Section 4 provides for an in-depth analysis of five open source solutions selected from the set of software reviewed from section 3. Finally, Section 5 concludes the paper.

## 2. DIGITAL INFORMATION PRESERVATION: MODELS, STANDARDS AND PROTOCOLS

We review OAIS and some other main standards, as well as, some protocols that are relevant for digital information preservation.

### 2.1 The OAIS Reference Model

The emergence of digital information preservation took a while. In 1994, a task force was created from the joint effort of two groups,the Commission on Preservation and Access (CPA) and the Research Libraries Group (RLG), both comprised of archivists and publishers. This task force studied the needed actions for ensuring long-term preservation and continued access to digital materials. Later, the Consultative Committee for Space Data Systems (CCSDS) was asked to define rules and methodologies for long term archival/storage of digital data generated from space missions. The result of this effort was the Open Archival Information System (OAIS) reference model. OAISis the first reference model on digital data preservation[15].Itbecame a standard for digital information preservation in January 2002 as ISO-STD 14721. In 2012, an updated version of this model was published (ISO 14721:2012) [14].This model focuses on providing a structure along with a lexicon of well-defined concepts and frameworks for any archive or system to be built with the purpose of preserving information and making it available for long-term use by a designated community or target group.

To be OAIS-aware, an information preservation solution needs to provide functionality to deal with ingestion, preservation and dissemination of archived digital materials. For this purpose, the OAIS reference model defines that at least the following six high-level services need to be provided by the archival and preservation solutions. They are: i) ingest; ii) archival storage; iii) data management; iv)preservation planning; v) access and vi) administration[14]. On top of that, OAIS defines an environment with three main roles: *management*, *producer* and *consumer*. Management is in charge of the system, while a producer is the entity that aims to preserve data in the archive preservation solution. Finally, consumers are the individuals/organizations that can access the preservation system to retrieve information.

Regarding the content to archive and preserve, the OAIS model is centred on the information package. The information package comprises the object to preserve, the needed metadata for long time preservation, the access permissions and how the whole data should be interpreted when accessed. Specifically, the OAIS defines three distinct information packages: i) Submission Information Package (SIP); ii) Archival Information Package (AIP) and iii) Dissemination Information Package (DIP)[14]. The SIP represents the source information which is inserted into the archival system by the producer entity. The AIP is the information that is actually archived, complemented with the metadata needed for a proper preservation and future accesses. The DIP represents the information provided to a consumer's request. Its format and content may adapt to the profile of the consumer. For instance, a content archived under a given encoding format, e.g. UTF16, may be delivered to consumers in another encoding format, such as, UTF8.

Figure 1presentsthe main services and the functional entities of the OAIS reference model. The rectangle-shaped boxes represent the high level services that need to be provided by an OAIS-oriented preservation solution. As can be seen, a SIP is first processed by the ingestion module. The ingestion procedure of a SIP yields an AIP to be kept in the archival storage and a set of metadata that feeds the data management service. The AIP is the crux of the information preservation system. It comprises the original information to preserve, as well as, the data needed to interpret the information. OAIS recommends fourtypesof metadata: i) descriptive (provided by the user), ii) technical (extracted by specific tools), iii) preservation (data from operations carried out during the preservation process, e.g. checking of file checksums), and iv) structural(defines relationships between files)[14].

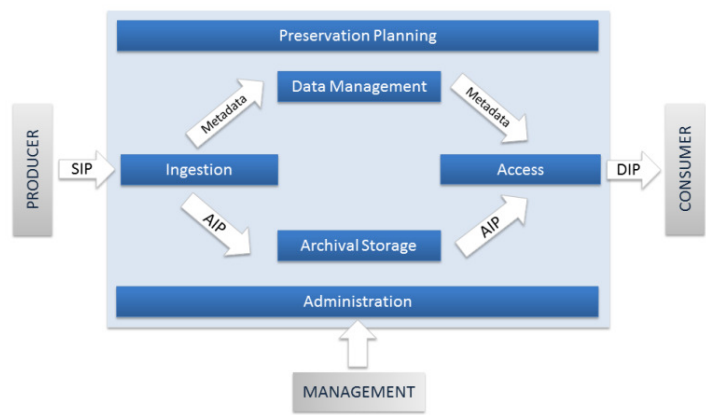

Figure 1. Open Archival Information System (OAIS) Reference Model (adapted from [15]).

When an access request is received by the archival and preservation system, the access service interacts with the data management and the archival storage services to produce the DIP as requested by the consumer. *Preservation planning*, shown at the top of

Figure 1, is a service transversal to the whole system. It represents the preservation strategy, dealing with external issues such as changes in technology (e.g., obsolescence of a given type of storage) or adjustment in the interaction with producers and consumers. Finally, *administration* is another transversal service. As the name implies, it deals with administrative issues. Specifically, it coordinatesto fulfil the needs of the other five main services, monitors the performance and manages the maintenance needs of the whole system.

On the matter of interoperability, the OAIS model defines three main categories: i) cooperating; ii) federated; and iii) shared functional areas [14]. Cooperating repositories provide for at least some compatibility at the SIP and DIP level. For instance, a DIP of one repository can be ingested, and thus can act as a SIP into collaborating repositories. Federated repositories aim to provide for integrated services, with a request for data (DIP) possibly filled by two or more distributed repositories. Finally, repositories can share resources needed to support functional activities such as ingestion, storage or data management, to name just a few.

**2.2 Main standardsand protocols**

Two main protocols for interoperability are the Open Archives Initiative Protocol for Metadata Harvesting (OAI-PMH)[16]and the Search/Retrieval via URL (SRU)[17]. OAI-PMHwas created by the Open Archives Initiative for repository interoperability. It consists of six verbs or services invoked over HTTP. The verbs/services are: i) GetRecord; ii) Identify; iii) ListIdentifiers; iv) ListMetadataFormats; v) ListRecords; vi) ListSets[16].The repositories can act as data providers, exposing structured data through the protocol or as service consumers making requests through the protocol to harvest metadata from the providers.SRU is a XML-based protocol to allow search queries over the internet. It uses the Contextual Query Language (CQL) standard[17], a syntax for representing queries to retrieve data from the repository and exposesthem in a structured form through XML.

Metadata standards define the main characteristics needed to describe digital objects, such as, videos, sounds, images, texts and web sites. The main standards are Dublin Core[18],

PREservation Metadata Implementation Strategies (PREMIS)[19], Machine Readable Cataloguing (MARC)[20], Encoded Archival Description (EAD)[21], Metadata Encoding and Transmission Standard (METS)[22]and the Metadata Object Description Schema (MODS)[22].

The Dublin Core standard was created in 1995 and is maintained by the Dublin Core Metadata Initiative. It comprises 15 properties with metadata vocabularies and technical specifications, which can describe a wide range of resources[18]. PREMIS, MARC, EAD and METS are all XML-based standards. PREMIS, developed by the Online Computer Library Center (OCLC) and Research Libraries Group (RLG), consists of a data dictionary, an XML schema and supporting documentation. MARC was developed by the American Library of Congress for cataloguing digital objects stored in a repository. METS isa part of MARCfor encoding descriptive, administrative and structural metadata about digital objects within a repository. MARC21 is the most recent version, while MARCXML is an extension of MARC21 with additional features for sharing and networked access of bibliographical information[23]. MODS is another MARC21 compatible XML for descriptive metadata. EAD is a descriptive XML-based standard aimed at describing the hierarchy structure of archival data. It bears some similarity with MARC, although EAD focuses on archives, while MARC is oriented towards bibliographical materials[21].

Others, more specific, standards and protocols considered in this paper are the Digital Item Declaration Language (DIDL) [24] for video content, the Metadata for Images in XML (MIX)[25] for still images, the Technical Metadata for Text (TextMD)[26] for text, the Advanced Encryption Standard (AES)[27]for encryption, the Document Mediated Delivery (DocMD) [28], AudioMD for audio[29], and VideoMD for video [29].It is important to note that regarding metadata standards, there is no consensus about which ones are the most important. As such, this study presents the ones implemented by each reviewed system.

## 3. COMPARISON OF DIGITAL INFORMATION PRESERVATION SOFTWARE

We compare eleven digital information preservation software solutions. The selection criteria include several aspects relevant to identify the most modern, flexible and reliable systems available to date. Next, we present the selection criteria.

The study is restricted to preservation software which are available under an open source license.The relevance to the field of the software solution is another important criterion. The relevance is estimated either by the size of the users' community or the size of the contributors' community or both. For this study, we selected the systems that have the broader communities supporting them. Complementing the community size, is the dynamicity of the project. Although it is not an exclusion factor for this study, most of surveyed projects are currently active, having released at least one version in the last 6 months. There are two exceptions: Archimède[30] and DAITSS [31]. Both were included due to their importance to the digital preservation field and target audience. Another important request for inclusion in this study is the adherence to state of the art digital preservation and metadata standards. The assessed systems implement the most relevant digital preservation and metadata standards to the field, namely, OAIS, OAI-PMH, SRU, Dublin Core, MARC, PREMIS, MARC and METS.

### 3.1. Open Source Digital Preservation Software Solutions

Table 1identifies the eleven software solutions surveyed in this study. It lists each solution's author and the classification given by authors for their software product.The table also identifies the studied version of the software, released year and the open source license under which it is available. The software solutions are presented in ascended alphabetical order of the project name.

Table 1. Main characteristics of surveyed software solutions.

| Software | Authors' classification | Latest Version/ Release Year | Licensing | Author(s) |
|---|---|---|---|---|
| Archimède[30] | Institutional Repository | 2.0 2006 | GPLv2 | Bibliothèque de l'Université Laval, France |
| Archivematica[32] | Digital Preservation System | 1.6 2016 | AGPLv3 | Artefactual, Canada |
| DAITSS [31] | Digital Preservation Repository System | 2.4.20 2014 | GPLv3 | Florida Center for Library Automation, Florida, USA |
| DSpace[33] | Institutional Repository | 6.0 2016 | BSD 3-Clause [34] | DuraSpace, Oregon, USA |
| EPrints[35] | Institutional Repository | 3.4 2016 | GPLv3 | University of Southampton, UK |
| Fedora [36] | Digital Repository | 4.6.0 2016 | Apache 2 | Fedora Community; DuraSpace Oregon, USA |
| Greenstone [37] | Digital Repository | 3.0.7 2016 | GPLv2 | New Zealand Digital Library Project University of Waikato New Zealand |
| Invenio[38] | Digital Library / Document Repository | 3.0 2016 | GPLv2 | CERN; DESY; EPFL; FNAL; SLAC |
| LOCKSS [39] | Library-led Digital Preservation System | 1.70.2-1 2016 | BSD | Stanford University California, USA |
| RODA [40] | Digital Repository | 1.2.0 2016 | LGPLv3 | Keep Solutions; University of Minho, Portugal |
| Xena[41] | Digital Preservation Software | 6.1.0 2016 | GPLv3 | The National Archives of Australia |

## 3.2. Main Features

We identify and describe the main features of the surveyed systems. The main identified features are as follows: i) digital preservation strategies; ii) authorization/authentication; iii) search capabilities; iv) previews; v) reporting capabilities; vi) support for multilingual; and vii) dynamism of the community of developers.

**Digital preservation strategies** focus on the strategies made available by each system to ensure long-term access, integrity and authenticity of stored data. **Authorization and authentication** features assess the existence of access control mechanisms and the ability to track users' action over data. **Advanced search** focuses on the capability of the software solution to allow for searches over stored data, letting users specify filters and properties of data. **Operating System (OS) Support** gives the information about the three main desktop operating systems: Linux, Windows and MacOS. **Previews** assess the ability of the preservation software to generate previews, thumbnails and other small excerpts of stored digital objects, saving users from the need to download full packages just to peek at their contents. **Reporting** evaluates the capability to produce simple/detailed reports with the preserved information. **Multilingual** assesses support for interaction in the user's native idiom. This feature is relevant for data dissemination purposes. Moreover, even if the information itself may solely exist in one language, it is still important that the repository software can support multiple idioms. Finally, **community of developers** focuses on whether there is a vast and dynamic set of developers involved with the software. This is especially important for open source software, as it may define the difference between failure and success.

Table 2lists the main features of the surveyed solutions. Besides the features presented in the table, all the surveyed solutions support authentication/authorization and allow for advanced search.

Table 2. Main features of the surveyed solutions.

| Software | Digital Preservation Strategies | OS Support | Previews/Reporting/Multilingual/ Community of Developers |
|---|---|---|---|
| Archimède | Encapsulation | Linux, Windows, MacOS | - / yes / yes / - |
| Archivematica | Migration, Emulation | Linux (Ubuntu) | - / - / - / - |
| DAITSS | Refreshment Migration | Linux, Windows, MacOS | - / - / - / - |
| DSpace | Encapsulation Federation | Linux, Windows, MacOS | - / yes / yes / yes |
| EPrints | Encapsulation | Linux, Windows, MacOS | yes / yes / yes / - |
| Fedora | Encapsulation Federation | Linux, Windows, MacOS | - / - / - / yes |
| Greenstone | Encapsulation | Linux, Windows, MacOS | yes / yes / yes / - |
| Invenio | Encapsulation | Linux | yes / yes / yes / yes |
| LOCKSS | Encapsulation Federation, Migration | Linux | - / - / - / - |
| RODA | Migration | Linux, Windows, MacOS | yes / yes / yes / yes |
| Xena | Migration | Linux, Windows, MacOS | - / - / - / - |

### 3.3 Preservation Standards / Metadata Standards Support

This section identifies the preservation and metadata standards supported by each software. The preservation standards are OAIS[14], OAI-PMH[16] (versions 1 and 2) and SRU[17]. The identified metadata standards are Dublin Core[18], MARC / MARC21 / MARCXML [42, 23], EAD [21], PREMIS [43], METS [44], MODS [45], DIDL [46], MIX [25], AES[27], DocMD[28]and TextMD[47].Table 3 shows the support for these standards given by each software. Specifically, in Table 3, a check sign means that the standard is supported. The inexistence of a check sign means that no indication of the standard's support was found during this study. Nonetheless, we cannot assert that any of the standards are unsupported, as many of the systems studied have flexible architectures, therefore supporting 3$^{rd}$ parties' plugins, capable of enabling those standards support.Note that Xena does not support any of the covered standards. Additionally, since only DAITSS natively supports the MIX, AES, TextMD, DocMD, audio and

28

International Journal of Computer Science & Engineering Survey (IJCSES) Vol.8, No.3, June 2017

video formats, these standards are not included in Table 3 to preserve space. For the same reason, the column MARC includes the standards MARC, MARC21 and MARCXML.

Table 3. Standards and protocols supported by each of the surveyed software solutions.

| Software | OAI | OAI-PMH v2 | SRU | Dublin Core | MARC | PREMIS | METS | MODS | DIDL |
|---|---|---|---|---|---|---|---|---|---|
| Archiméde | ✓ | ✓ | | ✓ | | | | | |
| Archivematica | ✓ | | | ✓ | | ✓ | ✓ | | |
| DAITSS | ✓ | | | | | ✓ | ✓ | | |
| DSpace | ✓ | ✓ | ✓ | ✓ | ✓ | | | ✓ | |
| EPrints | ✓ | | ✓ | ✓ | | | ✓ | ✓ | ✓ |
| Fedora | ✓ | | ✓ | ✓ | | ✓ | ✓ | | |
| Greenstone | ✓ | | ✓ | ✓ | ✓ | | ✓ | | |
| Invenio | ✓ | | ✓ | ✓ | ✓ | | | | |
| LOCKSS | | | | ✓ | | | | | |
| RODA | ✓ | ✓ | | ✓ | | ✓ | ✓ | | |
| Xena | | | | | | | | | |

Table 4. Supported file formats (part 1).

| Software | Image | Audio | Video |
|---|---|---|---|
| Archiméde | | | |
| Archivematica | BMP, GIF, JPG, JP2, PCT, PNG, PSD, TIFF, TGA, RAW | AC3, AIFF, MP3, WAV, WMA | AVI, FLV, MOV, MPEG-1, MPEG-2, MPEG-4, SWF, WMV |
| DAITSS | ✓ | ✓ | ✓ |
| D Space | ✓ | ✓ | ✓ |
| E Prints | JPG, JPEG, PNG, GIF, TIFF, AIFF | MP3 | MPEG |
| Fedora | ✓ | ✓ | ✓ |
| Greenstone | | | |
| Invenio | ✓ | ✓ | ✓ |
| LOCKSS | ✓ | ✓ | ✓ |
| RODA | TIFF, TIF, JPG, JPEG, PNG, BMP, GIF, ICO, XPM, TGA | WAV, MP3, MP4, FLAC, AIF, AIFF, OGG, WMA | MPG, MPEG, VOB, MPV2, MP2V, MP4, AVI, WMV, MOV, QT |
| Xena | BMP, GIF, PSD, PCX, RAS, XBM, XPM, PNG, TIFF | MP3, WAV, AIFF, OGG, FLAC | - |

## 3.4 Supported File Formats

We enumerate the file formats recognized by each system for data ingestion. Each system is capable of ingesting files of any type and storing them. However, only recognized file types allow

International Journal of Computer Science & Engineering Survey (IJCSES) Vol.8, No.3, June 2017

the systems to perform operations such as migration, normalization or other preservation operations specific to each file type on the ingestion phase.Table 4 and Table 5group file formats in eight sets:*Image*, *Audio, Video* (Table 4)and *Text,Applications, Vector, Email, and Other* (Table 5). These tables identify the file types recognized by each of the surveyed solutions. A check sign means that the file type is supported. On the contrary, the inexistence of a check sign means that no indication of support for the file formats was found during this study.

Table 5. Supported file formats (part 2).

| Software | Text | Applications | Vector | Email | Other |
|---|---|---|---|---|---|
| Archiméde | XML, HTML, PDF, RTF, TXT | MS Word, MS PowerPoint, OpenOffice | | | JavaBeans |
| Archivematica | TXT, PDF, RTF | DOCX, XLSX, PPTX, PPT, XLS, DOC, WPD | AI, EPS, SVG | PST, MailDir | X3F, 3FR, ARW, CR2, CRW, DCR, DNG, ERF, KDC, MRW, NEF, ORF, PEF, RAF |
| DAITSS | ✔ | ✔ | | | |
| D Space | ✔ | ✔ | | | |
| E Prints | PDF, HTML, ASCII, CSV, XML | MS Word | | | ZIP, GZ |
| Fedora | ✔ | | | | |
| Greenstone | TXT, PDF, HTML | MS Word, MS PowerPoint, MS Excel | | | |
| Invenio | ✔ | ✔ | | | |
| LOCKSS | ✔ | ✔ | | | |
| RODA | PDF, XML | DOC, DOCX, ODT, RTF, TXT, PPT, PPTX, ODP, XLS, XLSX, ODS | AI, CDR, DWG | EML, MSG | BIN |
| Xena | PDF, SQL, HTML, TXT | MS Office, MS Project MPP, Open Office, WordPerfect | | PST | MBX, ZIP, GZIP, TAR, GZ, JAR, WAR |

## 4. MOST RELEVANT OPEN SOURCE DIGITAL PRESERVATION SOLUTIONS

Five of the surveyed software solutions stand out as the most relevant ones for institutions looking to implement digital repositories. These solutions are: *RODA*, *DSpace*, *Fedora*, *Greenstone* and *EPrints*. These solutions are feature rich and have a broad community of users. They are, in most cases, the first option for digital library management and long-term preservation. All of these solutions implement the OAIS reference model with the exception of Greenstone. Still, Greenstone is included in this chapter due to its wide use by UNESCO countries[48].

We review each of the five projects, explaining their main purposes, providing some historical background and highlighting their main features. We also briefly reference the technologies used by each system. The information was collected on the projects' official websites, scientific publications and on official documentations (promotional leaflets, brochures, etc.).

### 4.1. RODA – Repository of Authentic Digital Objects

The Repository of Authentic Digital Objects (RODA) [49]is a digital repository licensed under the open source LGPLv3 license. It follows and provides functionality of the OAISmodel. It is developed by Keep Solutions, a Portuguese company, in cooperation with University of Minho, and its research community. RODA targets not only academic institutions that wish to build their own digital repository, but also museums, libraries or any other institutionwith similar needs[50].

The built-in preservation strategy of RODA is migration. It features all the steps this strategy encompasses, i.e., normalization, conversion, replication and preservation. It also supports other strategies, such as emulation or encapsulation through its extendibility and configuration capabilities. RODA supports several main standards and is capable of ingesting information, normalize objects for data preservation andallow to browse the repository. It also provides advanced search over the entire repository contents, previews of stored digital objects for text based objects, images, audio or video files and downloading the preserved information [40]. RODA has an advanced ingest workflow[40], supporting the ingest of new digital material, as well as, associated metadata in four distinct ways: i) online submission (self-archiving); ii) offline submission using a client application called "RODA-in" (offline self-archiving); iii) batch import by depositing SIPs via FTP or SMB/CIFS; and iv) integration with 3rd party document management software via invocation of SOAP Services or client API.

RODA has the following main features[40]:

- It provides for access control and permission management, with flexible configuration and tracking of user actions.
- It is vendor independent, being able to use the most convenient hardware and operating system.
- It is scalable through a service-oriented architecture that supports load balancingwith several servers.
- It has embedded preservation actions such as format conversions, normalization steps during ingest, checksum verifications, reporting actions, notification events and emails.
- It has extensibility capabilities and provides support for $3^{rd}$ party systems integration through the exposure of functionality via web services. This allows other systems to easily communicate with RODA and let them add more functionality to the system.
- It has multilingual support.

RODA is built on top of a plethora of technologies. The main ones areJAVA (programming language and implementation), Google Web Toolkit (user interface), OpenLDAP(Authentication), Fedora Linux (Data Services), ImageMagick, OpenOffice, GhostScript, JOD Converter, MEncoder, SoundConverter and gStreamer(migration and conversion), JHove/JHove2 andDROID (Digital Record Object Identification) for automatic validation and characterization.

### 4.2. DSpace

DSpace is a repository software built with data preservation in mind [51], and licensed under the BSD open source license. It enables easy and open access to all types of digital content including text, images, video and data sets. DSpace recognizes and manages a large number of file format and MIME types, such as the most common formats PDF, Word, JPEG, MPEG and TIFF files. Although out-of-the-box DSpace only recognizes common file formats, other formats can be managed through a simple file format registry. This way, it is possible to register any unrecognized format, so that it can be identified in the future [33].

DSpace is a full stack web application consisting of a database, storage manager and a front end.The web applications provide interfaces for administration, deposit, ingest, search, and access to assets stored and maintained on a file system or on similar storage system. This way, it is highly customizable and configurable through a web-based interface [52]. Additionally, DSpace provides for full import/export of the repository feature for disaster recovery.The system provides for two main preservation strategies, encapsulation and federation. The metadata, including access and configuration information, is stored in a relational database. Under the federation strategy, DSpace acts as a peer repository in a decentralized network of repositories. DSpace is cross platform, supporting Linux, MacOS and Windows.

The benefits from a large community of developers and contributors who keep evolving and improving its features make DSpace one of the most used solution for libraries, educational institutions, governments, non-profit and even commercial organizations. Originally, the project, developed by MIT Libraries and Hewlett-Packard (HP) Labs, had its first release in 2002. The community is currently under the control of DuraSpace,an independent not-for-profit organization formed in 2009 by merging Fedora Commons and DSpace. Since then, DuraSpaceinvests in open technologies that promote durable, persistent access to digital data. It collaborates with academic, scientific, cultural, and technology communities by supporting projects and creating services to help the preservation of the collective digital heritage [33].

DSpacehas the following main features:

- Configurable file storage, either local file system or cloud-based service.
- Configurable workflows laid on top of specific data model architecture.
- Configurable metadata schemas through the mapping or specification of new fields over the default Dublin Core structure.
- Configurable browse and search, as well as, full text search capabilities.
- Built-in authentication/authorization system that can be integrated with 3[rd] party authentication mechanisms.
- Multilingual support.

DSpaceaims for open standards compatibility. To this purpose, it supportsvarious standards, namely: OAIS, OAI-PMH, Dublin Core, OAI-ORE (Open Archives Initiative Object Reuse and Exchange), SWORD (Simple Web-service Offering Repository Deposit), WebDAV (Web-based Distributed Authoring and Versioning), OpenSearch, OpenURL, RSS (Really Simple Syndication), and ATOM.

The main technologies in use by DSpace areJAVA(programming language), Angular 2 (user interface), LDAP and Shibboleth (3[rd] party authentication), and as database engines, PostgreSQL and Oracle.

International Journal of Computer Science & Engineering Survey (IJCSES) Vol.8, No.3, June 2017

## 4.3. Fedora

Fedora is a repository software suite that provides management and dissemination of digital content. It is licensed under the Apache 2 open source license. It targets digital libraries and archives. Fedora features in the list of the most widely used repository software. It has an established user base of academic institutions, universities, libraries and government agencies. The software is conceived for both data access and preservation. Fedora is able to provide specialized access to very large and complex digital collections of historic and cultural materials, as well as, scientific data [36].

The project was born in 1997 at the Cornell University in Ithaca, New York under the name of Flexible Extensible Digital Object Repository Architecture. It later adopted the Fedora acronym as its official designation after being referred to by that name in a scientific article[53]. Besides being born clearly before the Fedora Linux distribution by Red Hat, some legal issues were raised about the software designation. However, both parties agreed to maintain the Fedora name associated to their projects, as long as there was a clear association with the digital repositories systems in one case and the open source computer operating system in the other.

The Fedora Repository is supported by a large community of developers, led by the Fedora Leadership Group and is under the stewardship of DuraSpace not-for-profit organization [36].

Fedora has a robust and scalable architecture that enables it to handle collections with millions of objects [36]. It ensures the longevity and durability of data by storing all information in files without any software dependency and allowing the rebuilding of the complete repository at any time. It adheres to open standards, providing services via RESTful APIs.It also implements semantic web capabilities by resorting to the SPARQL query language to query repositories[54]. It supports the definition of complex relations between the digital objects stored. In the latest release, federation capabilities were also added, allowing the software to act as a peer repository in a distributed network of digital preservation repositories [55]. Fedora also allows for an easy deployment, resorting to a WAR file (Web application ARchive).

The main features of the digital preservation software Fedora are:

- Advanced storage options for files and metadata with customizable file systems and databases.
- Authentication, authorization and access control through integration with 3$^{rd}$ party standards compliant authentication frameworks.
- Pluggable security authorization modules: role-based, XACML or Web Access Control.
- Extensibility through plug-in modules capable of providing OAI-PMH dissemination or SWORD deposit.
- Advanced search, indexing and discovery through 3$^{rd}$ party applications.
- Preservation services such as fixity checking, audit trail, versioning, backup and restore.
- Batch operations capabilities over a single repository to achieve better consistency and performance.
-

Regarding technologies, Fedora resorts to Java (programming language and implementation), LDAP and Shibboleth (3$^{rd}$ party authentication), and MySQL and PostgreSQLas database engines.

33

**4.4. Greenstone**

Greenstone is a software suite for building and distributing digital library collections, and is licensed under GNU GPLv2. It is aimed for educational institutions, universities, libraries, public service institutions and UNESCO partner communities who wish to build their own digital libraries, especially in developing countries. Greenstone provides a way of collecting and organizing digital collections, publish them on the web or act as a standalone application and store the information in any storage medium, either hard drives or any removable media. In spite of being a digital repository software, Greenstone does not follow the OAIS reference model or implement explicitly any data preservation strategy. Despite not being a digital preservation repository per se, Greenstone is included in this study due to the fact that it implements some key features for data dissemination. Greenstone is also relevant to the digital repository target audiences [37].

Greenstone is produced by the New Zealand Digital Library Project at the University of Waikato. It is developed and distributed in cooperation with UNESCO and the Human Info NGO in Belgium.Greenstone can be run as a web server, with full search capabilities and metadata-driven digital resources. Alternatively, it can be run on a non-networked environment as a standalone application, being installed on a computer or operating from removable media. Greenstone also has a server version for the Android platform with the digital library self-contained on an Android device. This might be particularly interesting for anyone who wishes to make a library available without having to assemble and configure a conventional web server.

The software has interoperability capabilities with other systems through the implementation of contemporary standards like OAI-PMH or METS for metadata. Due to its support of these protocols, Greenstone is capable of interchanging information with systems like DSpace. This allows it to export/import from DSpace any collection available within these formats [37].

Other main features of Greenstone are as follows:

- Authentication/authorization service through JAAS (Java Authentication and Authorization Service).
- Built-in metadata management
- Built-in advanced search with customization possibilities.
- Built-in librarian interface that can manage remote Greenstone installations.
- Multilingual support.

Greenstone supports the standards OAI-PMH, METS, Dublin Core (qualified and unqualified), and Bibliographic records as specified by RFC 1807 [56]. It also supports AGLS and NZGLS. AGLS (Australian Government Locator Service) is an extension to the Dublin Core [57], to improve the visibility, manageability and interoperability of governmental online services, while NZGLS (New Zealand Government Locator Service) is based on AGLS. Itis a metadata standard implemented and maintained by the Archives of New Zealand, with the goal of classifying and categorizing New Zealand's government agency information and services[58].

Technology-wise, Greenstone relies mostly on Java for implementation and user interface. The authentication and authorization is performed through JASS (Java Authentication and Authorization Service).

## 4.5. EPrints

EPrints[35]is a software package for building open access repositories, licensed under GNU GPLv3. EPrints is primarily used for institutional repositories and scientific journals as it provides open access, i.e., immediate online access to the full text of research articles within the repository. Its flexible configuration and web-based nature allow it to be also used as a repository for images, research data or audio archives. EPrints provides a set of ingest, preservation, dissemination and reporting services for institutions open access needs [59].

EPrints was created in 2000 as a result of the 1999 Santa Fé meeting, where the discussion for the creation of a communication protocol for digital repositories interoperability gave birth to the OAI-PMH protocol [59]. EPrints provides a stable yet flexible infrastructure on which institutions have been building their open access digital repositories. Examples includes governmental departments, universities, hospitals and non-profit organizations [60]. Through EPrints Services – a not-for-profit commercial services organization – academic and research institutions can benefit from training, as well as, aid on the development and hosting of repositories. The project has been developed at theUniversity of Southampton, School of Electronics and Computer Science. It encompasses developers, librarians and users.

The software is a full stack web application consisting of a database, storage manager and a customizable front end web interface. Besides the web-based application, EPrints provides a command-line interface. Both interfaces are based on the LAMP architecture, using the PERL programming language in substitution of LAMP's usual PHP language. EPrintsuses a plugin architecture for importing and exporting data, creating representation of objects appropriate for indexation of search engine and for user interface widgets. Plugins are developed in the PERL language.EPrints supports the ingestion of practically any type of file.

In addition to the above stated, EPrintshas the following features[35]:

- Advanced search with autocomplete features.
- Lightweight metadata collection with METS and MODS export plugins.
- Tagmechanism and collection-based methods to classify digital materials.
- Support for multiple idioms.

The following standards are available within EPrints: OAIS, OAI-PMH, SWORD, Dublin Core, METS, MODS (Metadata Object Description Schema) and DIDL (Digital Item Declaration Language)[24]. Regarding software, EPrints is based on PERL (programming language and implementation),HTML/CSS (user interface) and uses MySQL as its backend database server.

## 4.6. Other Digital RepositoriesSolutions

The digital repository solutions Archimède[30], Archivematica[61], DAITSS, Invenio[38], LOCKSS[39]and Xena[41]are also valid choices to deal with the digital preservations needs of an institution. However, only LOCKSS and Inveniohave a good level of acceptance by their target audiences. This may be due to a lack of features or to a lack of standards compliance of the other solutions. Some solutions address the challenge of preserving data from a standalone application approach. This is the case for Xena, a solution that may be suitable in some cases, but does not seem reliable in the long run. Other solutions, like LOCKSS, are trying to break ground by implementing new preservation strategies, federation, which may also be a drawback for users looking for a system with given proofs and reliability. Another limitation in most of these software solutions is that they were built as digital libraries management software, lacking features of general purpose digital repositories.

## 5. CONCLUSION

This survey reviewed the state of the art of digital preservation repository software, focusing exclusively on solutions available under an open source license.The claim for a raising awareness on the digital preservation of information importance and need is taking place, with many organizations elaborating plans to preserve their digital assets. A need once felt mainly by archivists and librarians, has now given place to a more generalized necessity. The software solutions have evolved from very specific to more general purpose repositories. They are able to ingest many different types of data and have important data recognition functionalities, much broader than the earlier solutions, which were mostly tailored for the needs of libraries and archives.

The most important contributor to open source-based digital preservation software is academia. Indeed, several academic institutions are actively developing their own digital library repositories, involving the scientific community and the community of users. This allows for testing real scenarios, to receive users' feedback and requests for new features, contributing for the enhancement and maturation of software solutions. Other open source projects exist outside academia, namely on governmental and also on private institutions.Some commercial models are also emerging, linked to open source solutions. In fact, a whole set of new companies are offering digital preservation professional services, consultancy and training, building their solutions on top of open source software and open standards and protocols. This contributes for a more stable and reliable digital preservation ecosystem.

Regarding standards and normalization, the continuous effort for the development and consolidation of the OAIS reference model, metadata and system interoperability standards has contributed to the quality of some of the digital preservation software solutions. New standards and preservation strategies are being developed and perfected. An example is federation, a preservation strategy which involves not only the most recent interoperability protocol OAI-PMH, but also the fine-tuned Dublin Core metadata standard for communication between systems. More project endorsers and support communities are joining this initiative because there is a consensus that decentralized repository networks are the future of digital preservation.

## ACKNOWLEDGEMENTS

Financial support provided in the scope of R&D Unit 50008, financed by the applicable financial framework (FCT/MEC through national funds and when applicable co-funded by FEDER – PT2020 partnership agreement).

## AUTHORS

**Carlos André Rosa** holds a B.Sc.(2007) in Computer Engineering, specialization in Information Systems from InstitutoPolitécnico de Leiria, Portugal. He is currently in the M.Sc. in Mobile Computing at the same institution. He works as a software engineer at the VOID software company. His interests include digital data preservation, user experience design, Internet of Things and electronics.

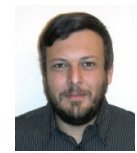

**Olga Craveiro** holds a Ph.D.(2016) in Information Science and Technology from the University of Coimbra, Portugal. She is professor at Department of Informatics Engineering of ESTG, Polytechnic Institute of Leiria, Portugal. Her main research areas are databases, databases administration, information retrieval and digital information preservation.

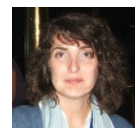

**Patricio Domingues** is with the Department of Informatics Engineering at ESTG - InstitutoPolitécnico de Leiria, Portugal. He holds a Ph.D. (2009) in Informatics Engineering from the University of Coimbra, Portugal. His research interests include multi-core and many-core systems, parallel computing and image, video processing, digital forensics and digital data preservation.

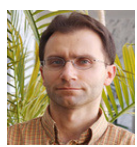